\documentclass[11pt]{article}
\usepackage{hyperref}
\usepackage{cite}
\usepackage{amsmath}

\usepackage{amsfonts}

\renewcommand{\title}[1]{%
    \bigskip%
    \begin{center}%
    \Large\bf #1%
    \end{center}%
    \vskip .2in}
    
\renewcommand{\author}[1]{%
    {\begin{center}
    #1
    \end{center}}}
\newcommand{\address}[1]{\vspace{-1.7em}\vspace{0pt}
    {\begin{center}
    \it #1 
    \end{center}}}
\begin{document}

\title{   Ward identities and gauge flow for M-theory in  ${\cal N}{=}3$  superspace}

 \author  { Sudhaker Upadhyay$\,^{\rm a}$ \footnote{Email: sudhakerupadhyay@gmail.com;  sudhaker@iitk.ac.in}} 
  \address{ $^{\rm a}${Department of Physics, Indian Institute of Technology Kanpur, Kanpur 208016, India}}

\begin{abstract}
We derive the BRST symmetry, Slavnov-Taylor identities and Nielsen identities for the ABJM theories
in  ${\cal N}{=}3$  harmonic superspace.
Further, the gauge dependence of one-particle irreducible amplitudes
in such superconformal Chern-Simons  theory is shown to be generated by a canonical
flow with respect to the extended Slavnov-Taylor
 identity, induced by the extended BRST transformations (including the BRST transformations of the gauge parameters).
\end{abstract}

\section{Introduction}
In the recent literature there has been a lot of excitement in search of the
superconformal Chern-Simons theory. The 
basic intention of doing so was to build a theory describing coincident M2-branes. 
M2 branes described by three-dimensional
superconformal field theories  have the structure of Chern-Simons-matter theory
with $ {\cal N}=6$ or ${\cal N}=8$ extended supersymmetry.
In fact, the Aharony, Bergman, Jafferis and Maldacena
(ABJM)  theory  which is the three-dimensional $ {\cal N}=6$ superconformal theory was constructed 
  to describe multiple M2 branes on the ${\mathbb C}^4/{\mathbb Z}_k$  orbifold \cite{ABJ}.
In ABJM
theory the  Chern-Simons gauge connections interact  with fermions
and scalars in bifundamental representations.  
  However,   Bagger, Lambert
and  Gustavsson (BLG) theory \cite{blg} is a superconformal theory which follows   ${\mathcal N}=8$ supersymmetry.
  Such three dimensional conformal field theories is also important in the sense that  they
 describe   conformal fixed points in condensed matter systems. From this point of view the highly supersymmetric versions are more solvable and, therefore, are  more interesting   models.
  
  It is desirable to have a superfield description of the ABJM models  with maximal
number  off-shell supersymmetries. As in other cases, such superfield
formulations are expected to bring to light geometric and quantum properties of the theory.
Here we are interested in harmonic superspace. The concept of harmonic superspace was developed   by  Galperin, Ivanov, Ogievetsky and Sokatchev  in  1980s \cite{X}. 
The $ {\cal N}=2$ harmonic superspace, is standard superspace augmented by the
two-dimensional sphere $S^2  \sim \textrm{SU} (2)/U(1)$. 
The $ {\cal N}=2$ harmonic superspace has  isospinor harmonics in addition to the usual one. By introduction of isospinor harmonics  it is 
possible to  $\textrm{SU}(2)$-covariantise the notion of
Grassmann
analyticity \cite{G4, G13}.
This helps enormously  to the adequate off-shell unconstrained formulations,
just like chirality \cite{F13}, the simplest kind of Grassmann analyticity \cite{G8}, is
a basis in $ {\cal N}= 1$ supersymmetry. Such analyticity represents to build
 an analytic subspace of harmonic superspace whose odd dimension
is half of that of the full superspace. Also a very similar analyticity underlies the $ {\cal N}=3$ 
gauge theory \cite{G5, G6}. The ABJM theory has been analysed, particularly, in harmonic superspace  in
 Refs. \cite{har,iva}.

  Apart from such investigations, the BRST quantization of  the superconformal Chern-Simons theories  was subject of interest 
in recent past \cite{sud0,sud1,sud2,sudd}. The BRST quantization materialize  the gauge 
conditions described by gauge parameters. On the formal side, it has been known since a long time that gauge
dependence of amplitudes can be studied algebraically through (generalized)
Nielsen identities \cite{nil,del,pg} having their origin in BRST symmetry. To be more precise  these identities can be derived by
extending the  BRST differential  to the gauge parameters.
The BRST variation of the gauge parameters is given by classical anticommuting variables paired
into a so-called BRST doublet.  Recently, the BRST quantization is analysed for the ABJM theory in harmonic superspace in covariant gauges
 \cite{mirf}. However, the Ward identities as well as flow of gauge parameters in such theories has not studied yet albeit the substantial 
progress made.  The  	
canonical flow in gauge parameters is studied recently for (non-conformal theory) Yang-Mills theory 
 \cite{qu}. 

In this paper we consider   ${\cal N}=3,  d{=}3$  harmonic superspace
and their algebra. Furthermore,
we analyse the superconformal ABJM theory in such superspace. Remarkably, we notice that
the ABJM theory in harmonic superspace follows the gauge invariance. However, 
according to standard quantization methods, we need to fix the extra gauge freedom 
associated with gauge symmetry. We have, therefore, fixed it here at quantum level
by adding suitable gauge-fixing as well as induced  (super)ghost terms to the classical
superconformal Chern-Simmons-matter parts. The resulting action 
remians invariant under fermionic rigid BRST symmetry. This BRST symmetry helps to compute the
Slavnov-Taylor identites  for the tree-level vertex functional. 
Further we extend the BRST symmetry by incorporating the variations of gauge parameters
which help us to demonstrate the extended Slavnov-Taylor identites as well as Nielsen identity. 
Additionally, we show the gauge dependence of one-particle irreducible amplitudes
in ABJM  theory in harmonic superspace to be generated by a canonical
flow with respect to the extended Slavnov-Taylor
 identity.

We organize the paper as following. In Sec. 2, we provide the general ${\cal N}=3$ harmonic 
superspace conventions
and setup. Also we review the ABJM theory in ${\cal N}=3$ harmonic superspace
with their gauge symmetry.  In section 3, we compute the BRST symmery along with various 
identities, namely, Slavnov-Taylor identities and Nielsen identity.
These identities helps us to study the behaviour of gauge parmaters.
Further, in section 4, we discuss the canonical flow of gauge parameters in ABJM theory in harmonic 
superspace.  In the last section  we draw concluding remarks.
\section{ ABJM Theory in ${\cal N}{=}3$ harmonic superspace}
In this section we mainly recapitulate the conventions and algebra followed 
by ${\cal N}{=}3$ harmonic superspace  \cite{ZK,Z3}. We also embed  the ABJM theory in this setup and discuss 
 their gauge symmetry \cite{har}. 
 \subsection{Harmonic superspace}
Let us start   by reviewing the ${\cal N}{=}3$, $d{=}3$ harmonic superspace as originally advocated in  \cite{ZK,Z3}
along with the field models \cite{har}. 
Here  ${\cal N}{=}3$
superspace is described by
the following real coordinates \footnote{The  notations are setted  as follows:  the Greek letters $\alpha,\beta,\ldots$ denote the
spinorial indices corresponding to the $\textrm{SO}(1,2)\simeq \textrm{SL}(2,R)$ Lorentz group.} 
\begin{equation}
z=(x^m,
\theta_\alpha^{ij}),\quad \overline{x^m}=x^m,\quad
 \overline{\theta_\alpha^{ij}}=\theta_{ij\alpha}.
\end{equation} 
 Now,  the covariant spinor
derivatives and supercharges are given by
\begin{equation}
D^{kj}_\alpha=\frac\partial{\partial\theta^\alpha_{kj}}
 +i\theta^{kj\,\beta}(\gamma^m)_{\alpha\beta}\frac{\partial}{\partial x^m},\quad
 Q^{kj}_\alpha=
 \frac\partial{\partial\theta^\alpha_{kj}}
 -i\theta^{kj\,\beta}(\gamma^m)_{\alpha\beta}\frac{\partial}{\partial x^m}\,.
\label{Q}
\end{equation}
In order to construct the harmonic superspace  parametrized by  the new bosonic coordinates
 (known as the harmonics in superspace) $u^\pm_i$ one should write a  matrix belonging to 
 the coset $\textrm{SU}(2)/{ U}(1)$ \cite{GIKOS,X}. 

Now, the  harmonic superspace can now be described by the following coordinates:
\begin{equation}
\zeta_A=(x^{\alpha\beta}_A,
\theta^{++}_\alpha, \theta^{--}_\alpha, \theta^{0}_\alpha, u^\pm_i), \end{equation} 
where
 \begin{equation}
x^{\alpha\beta}_A=(\gamma_m)^{\alpha\beta}x^m_A=x^{\alpha\beta}
+i(\theta^{++\alpha}\theta^{--\beta}+\theta^{++\beta}\theta^{--\alpha}),
\end{equation}
and $\theta^{++}_\alpha, \theta^{--}_\alpha, \theta^{0}_\alpha,$  are harmonic decompositions of the   anticommuting coordinates $\theta^{ij}_\alpha$, given by
\begin{eqnarray}
 (\theta^{++}_\alpha,\theta^{--}_\alpha,\theta^0_\alpha)=
(u^+_iu^+_j\theta^{ij}_\alpha,u^-_iu^-_j\theta^{ij}_\alpha,
u^+_iu^-_j\theta^{ij}_\alpha).
\end{eqnarray}
Now the  harmonic  derivatives in the above coordinates are 
\begin{eqnarray}
{\cal
D}^{++}&=&u_i^+ \frac{\partial}{\partial u_i^-}+2i\theta^{++ \alpha}\theta^{0 \beta}
 \partial^A_{\alpha\beta}
 +\theta^{++\alpha}\frac\partial{\partial\theta^{0 \alpha}}
 +2\theta^{0 \alpha}\frac\partial{\partial\theta^{--\alpha}},\nonumber\\
{\cal D}^{--}&=&u_i^- \frac{\partial}{\partial u_i^+}
 -2i\theta^{--\alpha}\theta^{0 \beta}\partial^A_{\alpha\beta}
 +\theta^{--\alpha}\frac\partial{\partial\theta^{0 \alpha}}
 +2\theta^{0 \alpha}\frac\partial{\partial\theta^{++ \alpha}},
\nonumber\\
{\cal D}^0&=&u_i^+ \frac{\partial}{\partial u_i^+} - u_i^- \frac{\partial}{\partial u_i^-}+2\theta^{++ \alpha}\frac\partial{\partial\theta^{++ \alpha}}
-2\theta^{--\alpha}\frac\partial{\partial\theta^{--\alpha}},  
\end{eqnarray}
and the harmonic decompositions of  spinor derivatives are  
 \begin{eqnarray}
&&D^{++}_\alpha=u^+_iu^+_jD^{ij}_\alpha = \frac{\partial}{\partial
\theta^{--\alpha}},\quad
D^{--}_\alpha =u^-_iu^-_jD^{ij}_\alpha=\frac\partial{\partial\theta^{++ \alpha}}
 +2i\theta^{--\beta}\partial^A_{\alpha\beta}, \nonumber\\
 &&
D^0_\alpha
=u^+_iu^-_jD^{ij}_\alpha= \frac12\frac\partial{\partial\theta^{0 \alpha}}
+i\theta^{0 \beta}(\gamma^m)_{\alpha\beta}\partial/\partial
x^m_A. \end{eqnarray}
 The algebra satisfied by these derivatives  are
 \begin{eqnarray} 
&& [{\cal D}^{++}, {\cal D}^{--}]={\cal D}^0, \ \{D^{++}_\alpha, D^{--}_\beta\}=2i\partial^A_{\alpha\beta}, \quad \{D^{0}_\alpha,
D^{0}_\beta\}=-i\partial^A_{\alpha\beta},\quad \{D^{\pm\pm}_\alpha, D^{0}_\beta\} = 0\,,
\nonumber\\
&&   
[{\cal D}^{\mp\mp}, D^{\pm\pm}_\alpha]=2D^0_\alpha, \quad [{\cal D}^{0},
D^{\pm\pm}_\alpha]=\pm 2D^{\pm\pm}_\alpha, \quad [{\cal
D}^{\pm\pm}, D^0_\alpha]=D^{\pm\pm}_\alpha.
\label{D-alg}
\end{eqnarray}
 
The full and  analytic integration measures are given conveniently by
\begin{eqnarray}
&&d^9z=-\frac{1}{16}d^3x_{A}
(D^{++})^2 (D^{--})^2(D^{0})^2,\ 
d\zeta^{(-4)}=\frac{1}{4} d^3x_{A}du  (D^{--})^2(D^{0})^2.
\end{eqnarray}
\subsection{ABJM theory}
In this section we sketch briefly the superconformal Chern-Simons-matter  theory with $ {\cal N}=6$ supersymmetry in harmonic superspace as in 
Ref. \cite{har}.
As the component content of the ABJM theory is given by four complex scalar fields and four complex spinor fields,    we first consider the two  gauge superfields for  ABJM theory in this  harmonic
superspace 
 $V^{++A}_{L\ \ B}$ and $V^{++A}_{R\ \ B}$ ($A, B= 1,2,...,N$) which are $N\times N$ matrices. These gauge superfields transform  under the gauge group 
 $U(N)_k$ and $U(N)_{-k}$,  respectively. 
To define the  ABJM theory we first write the gauge part of the action  as  \cite{har}
\begin{eqnarray}
S_{gauge} =
  \frac{ik}{16\pi}\mbox{Tr}\int d\zeta^{(-4)}[V_L^{++} D^{++\alpha} D^{++}_{\alpha}  V^{--}_L-V_R^{++}  D^{++\alpha} D^{++}_{\alpha}  V^{--}_R],
\end{eqnarray}
with non-analytic gauge superfields  
\begin{eqnarray}
 V^{--}_L&=&\sum_{n=1}^\infty (-1)^n \int du_1\ldots
du_n   \frac{V_L^{++}(z,u_1) V^{++}_L(z,u_2)\ldots  
V^{++}_L(z,u_n)}{(u^+u^+_1)(u^+_1u^+_2)\ldots (u^+_n u^+)}, \nonumber \\ 
 V^{--}_R&=& \sum_{n=1}^\infty (-1)^n \int du_1\ldots
du_n \frac{V_R^{++}(z,u_1) V^{++}_R(z,u_2)\ldots  
V^{++}_R(z,u_n)}{(u^+u^+_1)(u^+_1u^+_2)\ldots (u^+_n u^+)}.
\end{eqnarray}
We also define matter fields $q^{+}_a,\bar q_a^{+} (a=1,2)$, transform under the bifundamental representation 
of the group   $U(N)_k \times U(N)_{-k}$.
The gauge
invariant generalization of the hypermultiplet action is   \cite{har}
\begin{eqnarray}
 S_{M} [ q^{+}, \bar q^{+}] 
 &=&\mbox{Tr} \int d\zeta^{(-4)}\bar q_a^{+}  \nabla^{++}
   q^{+a},
\end{eqnarray}
where
the gauge covariant harmonic derivative $ \nabla^{++}={\cal  D}^{++} + V^{++}_L-V^{++}_R $.
Now, the classical action  for the  ABJM theory in harmonic superspace can now be given by
\begin{equation}
S_{ABJM} =S_{gauge}+S_{M},\label{act}
\end{equation}
which remains invariant under following gauge transformations  \cite{har}:
\begin{eqnarray}
&&\delta q^{+a} = \Lambda_L q^{+a}-q^{+a}\Lambda_R,\ \ \ \ \ \delta \bar q^{+a} = \Lambda_R\bar q^{+a}-\bar q^{+a}\Lambda_L,\nonumber\\
  &&\delta V_L^{++} =\nabla_L^{++} \Lambda_L =: -{\cal D}^{++} \Lambda_L -[V_L^{++}, \Lambda_L],\nonumber\\
  &&\delta V_R^{++} =\nabla_R^{++} \Lambda_R =: -{\cal D}^{++} \Lambda_R -[V_R^{++}, \Lambda_R],
\end{eqnarray}
where $\Lambda_L$ and $\Lambda_R$  are the gauge parameters. 
This model is also invariant under the following extra $\mathcal{N}= 3$ supersymmetric transformations  \cite{har}: 
\begin{eqnarray}
\delta_\epsilon q^{+a}&=& i\epsilon^{\alpha(ab)}\hat\nabla^0_\alpha  q^{+}_b\,,
\nonumber \\
\delta_\epsilon\bar q^{+}_a &=&i\epsilon^{\alpha}_{(ab)} \hat\nabla^0_\alpha  
\bar q^{+ b}\,, \nonumber  \\
\delta_\epsilon V^{++}_L&=&\frac{8\pi}k\epsilon^{\alpha(ab)}
 \theta^0_\alpha  q^+_a\bar q^+_b\,, \nonumber \\
\delta_\epsilon V^{++}_R &=&\frac{8\pi}k\epsilon^{\alpha(ab)}
 \theta^0_\alpha  \bar q^+_a  q^+_b\,,
\end{eqnarray}
where
\begin{eqnarray}
\hat\nabla^0_\alpha  q^{+}_b=D^0_\alpha q^+_b
 -\frac12D^{++}_\alpha
V^{--}_{L} q^+_b +\frac12 q^+_b D^{++}_\alpha
V^{--}_{R}
+\theta^{--}_\alpha(W^{++}_L  q^{+}_b -q^{+}_b  W^{++}_R ).
\end{eqnarray}
Thus, together with the original manifest $\mathcal{N} =3$ 
supersymmetry, this model has  $\mathcal{N} =6$ supersymmetry. 

 \section{BRST symmetry and Ward identities}
In this section we analyse the BRST symmetry of ABJM theory in ${\cal N}=3$ harmonic superspace.
As the ABJM model in harmonic superspace is gauge invariant it contains some spurious degrees 
 of freedom.  These extra degrees of 
freedom give rise to constraints in the canonical quantization  and 
divergences in the path integral  quantization \cite{ht}. To get rid of such
redundancy of degrees of freedom we restrict the gauge superfields to follow the certain
gauge-fixing conditions:
\begin{eqnarray}
 {\cal F}_L = {\cal D}^{++}V_L^{++} =0,\ \  {\cal F}_R = {\cal D}^{++}V_R^{++}=0.
 \end{eqnarray}
 The effect  of above gauge conditions can be incorporated at quantum level in the theory
by adding appropriate gauge-fixing terms   
in the classical   action (\ref{act}).
Here  the (linearized)  gauge-fixing terms are  \cite{G4}
 \begin{equation}
 S_{gf} =\int d\zeta^{(-4)} \mbox{Tr} \left[ -\alpha b_L{\cal F}_L   +\alpha b_R{\cal F}_R  \right],\label{ga}
 \end{equation}
where $b_L$ and $b_R$ are the multiplier superfields. According to the Faddeev-Popov quantization,
the   gauge-fixing terms induce the ghost terms in the functional integral.
Here the gauge-fixing terms (\ref{ga}) induce following ghost terms in the path integral:
 \begin{eqnarray}
 S_{gh}&=& \int d\zeta^{(-4)} \mbox{Tr} \left[ \alpha\bar c_L s {\cal F}_L -\alpha \bar c_R s {\cal F}_R 
 \right],\nonumber\\
 &=& \int d\zeta^{(-4)} \mbox{Tr} \left[ \alpha\bar c_L {{\cal D}}^{++} \nabla_L^{++}c_L  - \alpha\bar c_R{ {\cal D}}^{++} \nabla_R^{++}c_R  \right],
 \end{eqnarray}
 where $c_L, c_R$ and $\bar c_L, \bar c_R$ are ghost superfields and corresponding antighost superfields respectively
 and $s$ denote the BRST variation.  The BRST transformations  for the superfields
 are defined by 
 \begin{eqnarray}
&& s\ V_L^{++} =\nabla_L^{++}c_L,\ \ \ \ s\ V_R^{++} =\nabla_R^{++}c_R,\nonumber\\
&&s\ c_L=-\frac{1}{2}[c_L, c_L],\ \ \ s\ c_R=-\frac{1}{2}[c_R, c_R],\nonumber\\
&&s\ \bar c_L =b_L,\ \ \ \ \ \ \ \ \ \ \  \ \ \ s\ \bar c_R =b_R,\nonumber\\
&&s\    b_L=0,\ \ \ \ \ \ \ \ \ \ \ \ \  \ \ \ s\  b_R=0,\nonumber\\
&&s\     q^{+a}=c_Lq^{+a} -q^{+a}c_R,\   \ s\  \bar q^{+a}=c_R\bar q^{+a} -\bar q^{+a}c_L. \label{brs1}
 \end{eqnarray}
Under the above transformations  the effective action $S_{ABJM}+S_{gf}+S_{gh}$ is invariant. 

Now, we restricted to the pure gauge sectors of the theory where only  gauge and ghost fields have non-linear   BRST variations,
so their renormalization requires  the
introduction of external sources known as anti-superfields. These anti-superfields are coupled
to the BRST variation of the corresponding superfields as follows
 \begin{eqnarray}
 S_{af}=   \int d\zeta^{(-4)} \mbox{Tr} \left[ V_L^{++*}sV_L^{++}-c_L^*sc_L- V_R^{++*}sV_R^{++}+c_R^*sc_R \right],
 \end{eqnarray}
where U(1) charge of  the antifields $c_L^*$ and $c_R^*$ is $+4$.
These antifields are introduced to analyse theory at general ground. However, for
a linear covariant gauge such terms vanishes because the the antifields $c_L^*$ and $c_R^*$ 
do not exist for such case.
Here minus signs in front of the terms $c_L^*sc_L$ and $V_R^{++*}sV_R^{++}$ are introduced for consistency with the
Batalin-Vilkovisky (BV) bracket conventions.
Now, we are able to define the tree-level vertex functional as follows,
 \begin{eqnarray}
 \Sigma^{(0)} =S_{ABJM} +S_{gf}+S_{gh}+S_{af}.
 \end{eqnarray}
 This vertex functional obeys the following Slavnov-Taylor identity:
\begin{eqnarray}
{\cal S}(\Sigma^{(0)})&=& \int d\zeta^{(-4)} \mbox{Tr} \left[ \frac{\delta\Sigma^{(0)}}{\delta 
V_L^{++*}} \frac{\delta\Sigma^{(0)}}{\delta 
V_L^{++ }} -  \frac{\delta\Sigma^{(0)}}{\delta 
c_L^{ *}} \frac{\delta\Sigma^{(0)}}{\delta 
c_L } + b_L\frac{\delta\Sigma^{(0)}}{\delta 
\bar c_L }+   \frac{\delta\Sigma^{(0)}}{\delta 
V_R^{++*}} \frac{\delta\Sigma^{(0)}}{\delta 
V_R^{++ }} \right.\nonumber\\
& -&\left.  \frac{\delta\Sigma^{(0)}}{\delta 
c_R^{ *}} \frac{\delta\Sigma^{(0)}}{\delta 
c_R } + b_R\frac{\delta\Sigma^{(0)}}{\delta 
\bar c_R } \right]=0.
\end{eqnarray}
Notice that the linearity of the BRST transformation of the antighosts $\bar c_L, \bar c_R$  
do  not   require the introduciton of corresponding anti-superfields.
The above  identity   holds irrespectively of the particular form of
the gauge-fixing  chosen. For instance, some specific choices of the gauge 
(e.g. linear covariant gauges or the Landau gauge) further identities (the  auxiliary-field and the ghost equations) arise \cite{sor00}. 

Now, one can extend the BRST
symmetry to act on the gauge parameters  to derive an extended
 Slavnov-Taylor identity, leading to the  Nielsen identities \cite{x1,x2}. The
BRST variation of gauge parameters are
\begin{eqnarray}
s\ \lambda_i=\theta_i,\ \ s\ \theta_i =0,\ \ s\ \alpha =\theta, \ \ s\ \theta=0,\label{brs2}
\end{eqnarray}
where $\lambda_i, \theta_i, \alpha$ and $\theta$ are the (arbitrary) gauge parameters.
So, the extended BRST transformations can now be given by expressions (\ref{brs1}) and (\ref{brs2}) collectively. Under this extended BRST transformation the gauge-fixing fermion introduces
additional terms into the BRST exact parts
\begin{eqnarray}
S_{gf}+S_{gh} &=&s\int d\zeta^{(-4)} \mbox{Tr}\left[ -\alpha\bar c _L {\cal F}_L +\alpha \bar{c}_R {\cal F}_R \right],\nonumber\\
&=&\int d\zeta^{(-4)} \mbox{Tr}\left[  -\alpha b_L{\cal F}_L +\alpha
\bar c_Ls{\cal F}_L
-\theta \bar c_L{\cal F}_L +\alpha\bar c_L\left(\frac{\partial {\cal F}_L}{\partial\lambda_i} \theta_i
+ \frac{\partial {\cal F}_L}{\partial\alpha} \theta \right) \right.\nonumber\\
&+&\left. \alpha b_R{\cal F}_R -\alpha
\bar c_Rs{\cal F}_R
+\theta \bar c_R{\cal F}_R -\alpha\bar c_R\left(\frac{\partial {\cal F}_R}{\partial\lambda_i} \theta_i
+ \frac{\partial {\cal F}_R}{\partial\alpha} \theta \right)   \right].
\end{eqnarray}
Therefore, the tree-level classical action satisfies the
following extended Slavnov-Taylor identity:
\begin{eqnarray}
\tilde {\cal S} (\Sigma ^{(0)}) =\sum_i \theta_i\frac{\partial\Sigma ^{(0)}}{\partial\lambda_i}
+\theta \frac{\partial\Sigma ^{(0)}}{\partial\alpha} +{\cal S} (\Sigma ^{(0)})=0.\label{ext}
\end{eqnarray}
In case of non-anomalous theories this equation holds upto the full vertex functional $\Sigma$:
\begin{eqnarray}
\tilde {\cal S} (\Sigma ) =\sum_i \theta_i\frac{\partial\Sigma }{\partial\lambda_i}
+\theta \frac{\partial\Sigma}{\partial\alpha} +{\cal S} (\Sigma)=0.
\end{eqnarray}
Now, we compute the Nielsen identity by taking derivative of $\Sigma$ with respect to $\theta$
and then setting $\theta, \theta_i$ equal to zero
\begin{eqnarray}
\left . \frac{\partial\Sigma}{\partial\alpha}\right |_{\theta=\theta_i=0} &=&-\left.\int 
d\zeta^{(-4)} \mbox{Tr}\left[\frac{\delta^2\Sigma}{\partial\theta\delta V_L^{++*}}\frac{\delta\Sigma}
{\delta V_L^{++}} - \frac{\delta\Sigma}{\delta V_L^{++*}} \frac{\delta^2\Sigma}{\partial\theta\delta 
V_L^{++}} - \frac{\delta^2\Sigma}{\partial\theta\delta c_L^{*}}\frac{\delta\Sigma}{\delta c_L} 
-\frac{\delta\Sigma}{\delta c_L^*}\frac{\delta^2\Sigma}{\partial\theta\delta c_L}
\right.\right.\nonumber\\
&+&\left.\left. b_L\frac{\delta^2\Sigma}{\partial\theta\delta\bar c_L} +\frac{\delta^2\Sigma}
{\partial\theta\delta V_R^{++*}}\frac{\delta\Sigma}{\delta V_R^{++}} - \frac{\delta\Sigma}{\delta 
V_R^{++*}} \frac{\delta^2\Sigma}{\partial\theta\delta V_R^{++}} - \frac{\delta^2\Sigma}
{\partial\theta\delta c_R^{*}}\frac{\delta\Sigma}{\delta c_R} -\frac{\delta\Sigma}{\delta 
c_R^*}\frac{\delta^2\Sigma}{\partial\theta\delta c_R}
\right.\right.\nonumber\\
&+&\left.\left. b_R\frac{\delta^2\Sigma}{\partial\theta\delta\bar c_R} 
\right] \right |_{\theta=\theta_i=0}.
\end{eqnarray}
In the same fashion we can get an expression for the derivative of $\Sigma$ with respect to $\lambda_i$ by
taking the
derivative of the extended ST identity with respect to $\theta_i$.

The quantum action principle (QAP) \cite{sor00} describes  the structure of the ward identities
at the quantum level. For QAP is  applicable to the theories which are local, Lorentz invariant
and power counting renormalizable.  To prove the renormalizability of the ABJM theory in harmonic superspace we have to show stability. For that we split theeffective action at first order 
in loop expansion in to two parts: a finite part and a divergent part
\begin{eqnarray}
\Sigma^{(1)}=\Sigma^{(1)}_{\mbox{fin}}+\Sigma^{(1)}_{\mbox{div}}.
\end{eqnarray}
Due to linearity of Slavnov-Taylor identity ${\cal S}\Sigma^{(1)}_{\mbox{div}}=0$.
Now the divergence occuring in the quantum level can be reabsorbed by introduction of local counter 
terms ontained by redefining the fields. In this way, we are able to
prove the algebraic renormalization of the ABJM theory in harmonic superspace. 
\section{Canonical flow of gauge parameters}
In this section we analyse the canonical flow of gauge parameters.
Let us begin the section by defining the   antibracket (BV bracket)  as follows
\begin{eqnarray}
\{X, Y\} =\int d\zeta^{(-4)}\mbox{Tr} \sum_\phi \left[(-1)^{\epsilon_\phi(\epsilon_X +1)} 
\frac{\delta_l 
X}{\delta \phi} \frac{\delta_l Y}{\delta\phi^*} -(-1)^{\epsilon_{\phi^*} (\epsilon_X +1)}\frac{\delta 
_l X}{\delta \phi^*} \frac{\delta_l Y}{\delta\phi} \right].
\end{eqnarray}
where collective superfield $\phi\equiv (V_L^{++}, c_L, \bar c_L, b_L, V_R^{++}, c_R, \bar c_R, b_R )$ and collective anti-superfields
$\phi^*\equiv (V_L^{++*}, c_L^*, \bar c_L^*, b_L^*, V_R^{++*}, c_R^*, \bar c_R^*, b_R^* )$.
Here $\epsilon_\phi$ and $\epsilon_{\phi^*}$ denote the statistics of the superfields $\phi$ and
the anti-superfields $\phi^*$. However $\epsilon_X $ refers the statistics of the functional $X$.

With the help of above antibrackets the extended Slavnov-Taylor identity (\ref{ext}) is 
writte by
\begin{eqnarray}
\tilde {\cal S} (\Sigma ) =\sum_i \theta_i\frac{\partial\Sigma }{\partial\lambda_i}
+\theta \frac{\partial\Sigma}{\partial\alpha} + \frac{1}{2}\{\Sigma, \Sigma\}=0.
\end{eqnarray}
By taking a derivative with respect to $\theta$  we get
\begin{eqnarray}
\left . \frac{\partial\Sigma}{\partial\alpha}\right |_{\theta=\theta_i=0} &=& -\left .\left\{ \frac{\partial\Sigma}{\partial\theta}, \Sigma\right\}\right |_{\theta=\theta_i=0}.\label{con}
\end{eqnarray}
Here we note that the argument goes in the same way if
one takes a derivative with respect to $\theta_i$. The expression (\ref{con}) shows that the derivative of the vertex functional with respect to $\alpha$ is
obtained by a canonical transformtion (with respect to the antibracket) induced by
the generating functional 	$\frac{\partial\Sigma}{\partial\theta}$. One cannot solve  (\ref{con}) by simple exponentiation because the RHS  in general depends on $\alpha$  and therefore 
one needs   to make recourse to a Lie series.
To achieve this goal, we introduce  the following operator:
\begin{eqnarray}
\Delta_\Psi =\{\cdot, \Psi \} + \frac{\partial}{\partial \alpha}.
\end{eqnarray}
Now in terms of the Lie series the vertex functional $\Sigma$ is given by  
\begin{equation}
\Sigma =\sum_{n\geq 0} \frac{1}{n!} \alpha^n \left[\Delta_\Psi^n\Sigma^{(0)}\right]_{\alpha =0},
\label{li}
\end{equation}
where $\Sigma^{(0)}$ refers the vertex functional at $\alpha = 0$.
Here we  remark that the above equation holds irrespectively of the form of the  gauge-fixing (and in particular is independent of the existence of a
auxiliary field equation and of a ghost equation, validating the stability of the gauge-fixing
in certain  cases).

Now, for illustration purpose, we choose the 
Lorentz covariant gauges 
\begin{eqnarray}
{\cal F}_L ={\cal D}^{++}V_L^{++},\ \ {\cal F}_R ={\cal D}^{++}V_R^{++}.
\end{eqnarray}
For these particular choices, the extended BRST-exact terms are given by 
\begin{eqnarray}
S_{gf}+S_{gh}
&=&\int d\zeta^{(-4)} \mbox{Tr}\left[  -\alpha b_L{\cal D}^{++}V_L^{++} +\alpha
\bar c_L {\cal D}^{++}\nabla^{++}_L c_L
- \theta \bar c_L {\cal D}^{++}V_L^{++}   \right.\nonumber\\
&+&\left. \alpha b_R{\cal D}^{++}V_R^{++} -\alpha
\bar c_R {\cal D}^{++}\nabla^{++}_R c_R
  + \theta \bar c_R {\cal D}^{++}V_R^{++}  \right].\label{gg}
\end{eqnarray}
For these  gauges,  the auxiliary superfield equation and the superghost equation hold:
\begin{eqnarray}
&&\frac{\delta \Sigma}{\delta b_L} =  - \alpha {\cal D}^{++}V_L^{++},\ \ \frac{\delta \Sigma}{\delta b_R} =  - \alpha{\cal D}^{++}V_R^{++},\nonumber\\
&&\frac{\delta \Sigma}{\delta \bar c_L}
= \alpha{\cal D}^{++} \frac{\delta \Sigma}{\delta V_L^{++*}} +\theta {\cal D}^{++}V_L^{++},\ \ \frac{\delta \Sigma}{\delta \bar c_R}
= \alpha{\cal D}^{++} \frac{\delta \Sigma}{\delta V_R^{++*}} - \theta {\cal D}^{++}V_R^{++},
\end{eqnarray}
where the first two of the above equations imply that the   auxiliary superfields-dependence are confined at tree level. However, the last two of the above equations in turn imply that at higher orders $(n \geq 1)$ $\Sigma$
  can depend on  $\bar c_L, \bar c_R$ only through the combinations
\begin{eqnarray}
\tilde{V}_L^{++*} =V_L^{++*} -{\cal D}^{++}\bar c_L,\ \ \tilde{V}_R^{++*} =V_R^{++*} -{\cal D}^{++}\bar c_R.
\end{eqnarray}
Imploying the above redefinitions  we define the 
  reduced functional as follows,
\begin{eqnarray}
\tilde{\Sigma} =\Sigma -\int  d\zeta^{(-4)} \mbox{Tr}\left[   -\alpha b_L {\cal D}^{++}V_L^{++}  +
\alpha b_R {\cal D}^{++}V_R^{++} \right].
\end{eqnarray}
With such introduction of the  reduced functional the antibrackets can be restricted 
to the variables $(V_L^{++},V_L^{++*})$, $(V_R^{++},V_R^{++*})$, $(c_L, c_L^{*})$ and $(c_R, c_R^{*})$ and therefore
flow equation reads
\begin{eqnarray}
\left . \frac{\partial\tilde\Sigma}{\partial\alpha}\right |_{\theta=\theta_i=0} &=&  -\int 
d\zeta^{(-4)} \mbox{Tr}\left[ \frac{\delta \Psi}{\delta V_L^{++}}  \frac{\delta \tilde\Sigma}{\delta \tilde V_L^{++*}} +\frac{\delta \Psi}{\delta \tilde V_L^{++*}}  \frac{\delta \tilde\Sigma} {\delta 
V_L^{++}} - \frac{\delta \Psi}{\delta c_L}  \frac{\delta \tilde\Sigma}{\delta c_L^{*}} -
 \frac{\delta \Psi}{\delta c_L^{*}}  \frac{\delta \tilde\Sigma}{\delta c_L}\right. \nonumber\\
&+&\left.  \frac{\delta \Psi}{\delta V_R^{++}}  \frac{\delta \tilde\Sigma}{\delta \tilde V_R^{++*}} +\frac{\delta \Psi}{\delta \tilde V_R^{++*}}  \frac{\delta \tilde\Sigma} {\delta 
V_R^{++}} - \frac{\delta \Psi}{\delta c_R}  \frac{\delta \tilde\Sigma}{\delta c_R^{*}} -
 \frac{\delta \Psi}{\delta c_R^{*}}  \frac{\delta \tilde\Sigma}{\delta c_R}\right].
\end{eqnarray}
Now, the
Lie operator $\Delta_\Psi$  is given by
\begin{eqnarray}
\Delta_\Psi (X) &=&\{X, \Psi\} +\frac{\partial X}{\partial \alpha},\nonumber\\
&=&\int 
d\zeta^{(-4)} \mbox{Tr}\left[ \frac{\delta X}{\delta V_L^{++}}  \frac{\delta \Psi }{\delta \tilde V_L^{++*}} +\frac{\delta X}{\delta \tilde V_L^{++*}}  \frac{\delta \Psi} {\delta 
V_L^{++}} - \frac{\delta X}{\delta c_L}  \frac{\delta \Psi}{\delta c_L^{*}} -
 \frac{\delta X}{\delta c_L^{*}}  \frac{\delta \Psi}{\delta c_L}\right. \nonumber\\
&+&\left.  \frac{\delta X}{\delta V_R^{++}}  \frac{\delta \Psi}{\delta \tilde V_R^{++*}} +\frac{\delta X}{\delta \tilde V_R^{++*}}  \frac{\delta \Psi} {\delta 
V_R^{++}} - \frac{\delta X}{\delta c_R}  \frac{\delta \Psi}{\delta c_R^{*}} -
 \frac{\delta X}{\delta c_R^{*}}  \frac{\delta \Psi}{\delta c_R}\right] +\frac{\partial X}{\partial \alpha}.
\end{eqnarray}
By expression (\ref{gg}) we see that at tree-level gauge-fixing fermion	 reduces to
\begin{eqnarray}
\Psi = O(\hbar).
\end{eqnarray}
The Lie series of the vertex functional $\Sigma$  given in (\ref{li}) then allows one to express the coefficients of
the $\alpha$-expansion of one-point irreducible amplitudes in the Lorentz-covariant gauge in terms
of one-point irreducible Landau gauge amplitudes plus an $\alpha$-dependent contribution, arising
from the 	 gauge-fixing fermion $\Psi$.
\section{Conclusion}
 In this paper we have  considered the
 ABJM theory in ${\cal N}=3$ harmonic superspace. 
 The field content of the ABJM model is given by four complex
scalar and spinor fields which live in the bifundamental representation of the $U(N)\times U(N)$
gauge group.  Besides their general setup   the
 gauge symmetry of the theory are also presented. As we know a gauge theory can't be quantized without getting rid of
 spurious gauge freedom. Therefore, we have chosen the covariant gauges in order to fix the 
 the gauge freedom. We have achieved this at quantum level by adding suitable gauge-fixing and the ghost actions to the classical action. Resulting effective action admits
 the BRST symmetry. Furthermore, we have computed the Ward identites for such theory in ${\cal N}=3$ superspace. With the help of this set  of Ward identites  we have 
 established the algebraic renormalizability of the ABJM theory in harmonic superspace.
  To see the behaviour of gauge parameters we have derived the Nielsen identities
 by extending the quantum action.  Such an extended quantum action remains symmetric under larger 
 set of BRST transformations.   This identities will be helpful to demonstrate the gauge independence of the gauge self-energy, and of the matter mass shell in case of ABJM theory.
 It was demonstrated in the case of ABJM theory that the 
identities lead to results complementary to those of the usual Ward identities. As with
the Ward identities, the Nielsen identities offer possibilities to check one’s calculations,
however, they also allow us to see where physical meaning may
be found in apparently gauge dependent Green’s functions.
 
 Further, the existence of a canonical flow in the space of gauge parameters and the
related solution in terms of a Lie series provide a way to analyse
the results within an algebraic framework. As   the generating functional of the canonical flow 
depends on the
gauge parameters, we are unable to get the full solution by a naive exponentiation.
The results  is bound to hold even
beyond perturbation theory (as far as the ST identity is valid).
Such a solution can be expressed only through an appropriate Lie series.
Knowing such a Lie series makes the comparison between computations carried out in different gauges
for ABJM theory in harmonic superspace
easy.
The relations
derived here are  particularly in the perturbaive sector. However, it will be interesting to
analyse such discussion  in the non-perturbative regime.
The two point function for gauge connection,  under the assumption that analyticity
in the gauge parameter around α = 1 holds, a closed formula interpolating between the Landau and the
other suitable covariant gauge can be obtained.

\end{document}